\documentclass[lettersize,journal]{IEEEtran}
\usepackage{amsmath,amsfonts}
\usepackage{algorithmic}
\usepackage{algorithm}
\usepackage{array}
\usepackage{textcomp}
\usepackage{subcaption}
\usepackage{url}
\usepackage{verbatim}
\usepackage{graphicx}
\usepackage{cite}
\usepackage{xcolor}

\newcommand{\hdel}[1]{}

\begin{document}

\title{Generative AI-Empowered Semantic Twin Channel Model for ISAC}

\author{Yi Chen,~\IEEEmembership{Member,~IEEE,} Yatao Hu, Ming Li,~\IEEEmembership{Senior Member,~IEEE,} and Chong Han,~\IEEEmembership{Senior Member,~IEEE}
\thanks{Yi Chen, Yatao Hu, and Ming Li are with the School of Information and Communications Engineering, Dalian University of Technology, Dalian, China (email: chenyi@dlut.edu.cn; yataohu@mail.dlut.edu.cn; mli@dlut.edu.cn).}
\thanks{Chong Han is with the Terahertz Wireless Communications Laboratory, Shanghai Jiao Tong University, Shanghai, China (email: chong.han@sjtu.edu.cn).}
}
\markboth{}%
{Shell \MakeLowercase{\textit{et al.}}: A Sample Article Using IEEEtran.cls for IEEE Journals}
\maketitle

\begin{abstract}
Integrated sensing and communication (ISAC) increasingly exposes a gap in today’s channel modeling. Efficient statistical models focus on coarse communication-centric metrics, and therefore miss the weak but critical multipath signatures for sensing, whereas deterministic models are computationally inefficient to scale for system-level ISAC  evaluation. This gap calls for a unifying abstraction that can couple what the environment means for sensing with how the channel behaves for communication, namely, environmental semantics. This article clarifies the meaning and essentiality of environmental semantics in ISAC channel modeling and establishes how semantics is connected to observable channel structures across multiple semantic levels. Based on this perspective, a semantics-oriented channel modeling principle was advocated, which preserves environmental semantics while abstracting unnecessary detail to balance accuracy and complexity. Then, a generative AI–empowered semantic twin channel model (STCM) was introduced to generate a family of physically plausible channel realizations representative of a semantic condition. Case studies further show semantic consistency under challenging multi-view settings, suggesting a practical path to controllable simulation, dataset generation, and reproducible ISAC benchmarking toward future design and standardization.
\end{abstract}


\section{Introduction}
\IEEEPARstart{I}{ntegrated} sensing and communication (ISAC) is identified by IMT-2030 as a key enabler for next-generation wireless networks, as it allows existing communication infrastructure to provide both connectivity and environment awareness with shared spectrum, hardware, and waveforms~\cite{itu}. Realizing this vision critically relies on accurate and reproducible ISAC channel models. As the foundation of ISAC network design, channel modeling underpins waveform and transceiver design, sensing algorithm development, and system-level performance evaluation~\cite{ywf,zjh-ISAC}. Reflecting this urgency, the 3GPP 102nd TSG RAN plenary meeting in December 2023 highlighted ISAC channel modeling as a primary research objective for the 6G standardization cycles.

Beyond the well-known extension from bi-static to mono-static deployment, ISAC fundamentally changes the modeling objective and consequently, the information a channel model must preserve. Communication channel models are designed to reproduce communication-centric metrics, e.g., path loss and delay spread, closely tied to communication performance. Those metrics are dominated by a few strong components, e.g., the LoS path and reflections from buildings. These metrics are inherently insensitive to sensing-relevant multipaths produced by sensing targets, which are often orders of magnitude weak and masked by the dominant components and noise. Therefore, ISAC channel modeling cannot directly follow the modeling principle of communication channels, leading to an irrecoverable loss of the information required by sensing tasks~\cite{gk-ISAC}. The objective for ISAC channel modeling is thus to retain sensing-relevant channel structures while maintaining channel characteristics governing communication performance.

Environmental semantics, a high-level abstraction of the propagation environment, provides a principled way to bridge the objective mismatch in ISAC channel modeling. It captures the meaningful information about objects, scenes, and events relevant to a given sensing or communication task, rather than exhaustively specifying all geometric and physical details~\cite{channel-semantics,zzy-cs}, while tightly coupled to the channel characteristics that govern communication performance. In this sense, semantics provides a common language that connects sensing-oriented channel meaning with communication-oriented channel behavior, and motivates a semantics-oriented ISAC channel modeling theory and methodology. Operationally, this requires learning a semantics-to-channel mapping under electromagnetic propagation constraints, which is inherently high-dimensional, highly non-linear, and non-Gaussian. Generative AI is well suited for such conditional mappings, as it can learn from distributions from data and represent complex dependencies. In the sequel, we leverage this capability in a semantic twin channel model (STCM) for ISAC. Unlike a digital twin channel model aiming to replicate a particular environment in detail~\cite{zjh-ChannelGPT}, a STCM is guided by environmental semantics and preserves the critical information required by ISAC.
\begin{figure*}[ht!]
\centering
\includegraphics[width=0.8\textwidth]{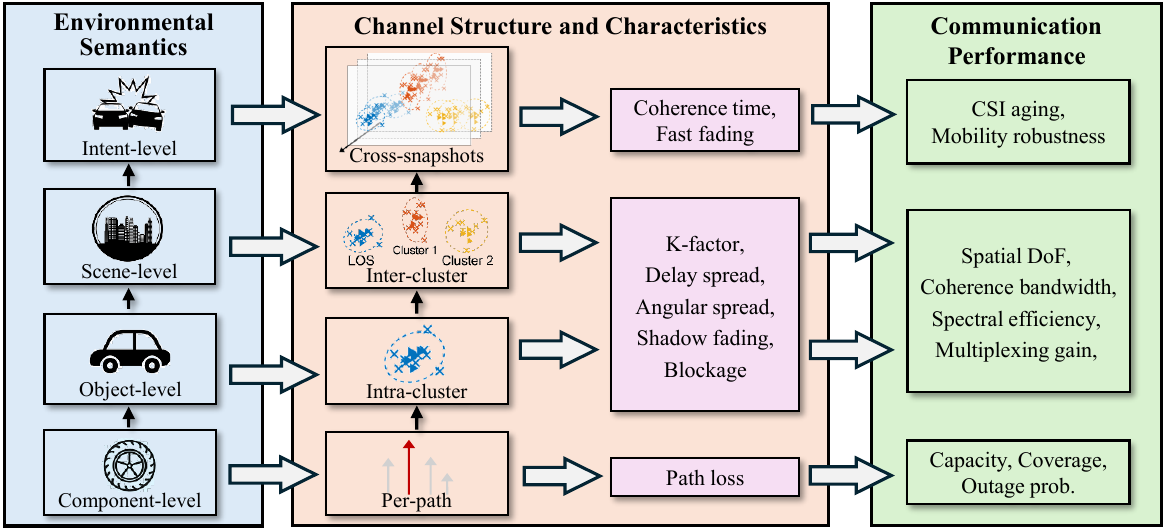}
\caption{Hierarchical Semantics-to-Channel Mapping.}
\label{fig:mapping}
\end{figure*}
\par This article advocates environmental semantics as the organizing principle for ISAC channel modeling. First, we define environmental semantics and its hierarchical mapping to channel structures. Then, we clarify the essentiality and requirement for an semantics-oriented ISAC channel model. Next, we propose a generative AI–empowered STCM that produces physically plausible channel realizations conditioned on environmental semantics. This channel model is embedded with a hybrid channel synthesizer that anchors the generated channels to fundamental propagation constraints and ensures physical plausibility. Finally, case-study validations to the proposed STCM are provided, followed by challenges and future directions toward practical adoption and standardization.

\section{Environment Semantics in ISAC Channel Modeling}
\subsection{Definition and Key Properties}
In a ISAC system, sensing ultimately aims to infer environmental semantics from the received signals, i.e., the meaningful information about objects, events, and scenes.
Environmental semantics refers to a high-level abstraction of the propagation environment that captures what is relevant to a given sensing or communication task, rather than an exhaustive specification of all geometric and physical details.
 Thus it can be instantiated through labels, attributes, and relations that admit unambiguous operational meaning. The corresponding semantic knowledge space therefore forms a subset or a low-dimensional projection of the full physical information space.

Environmental semantics suitable for ISAC channel modeling have three key properties. (1) \textit{Plausibility}: the specified entities and phenomena should exist in realistic environments, which ensures that channels generated under a given semantic condition are physically realizable and couldn't mislead algorithm development and performance evaluation.
(2) \textit{Diversity}: the same semantic element may correspond to a family of realizations with different geometries, materials, and micro-motions, which can induce different multipath structures while preserving the same high-level meaning. 
(3) \textit{Distinguishability}: semantic elements should not collapse into indistinguishable channel signatures. Practically, this can be formalized by endowing the semantic space with a metric and requiring a minimum separability margin. 

\subsection{Hierarchical Semantics-to-Channel Mapping}
Although the specific semantic elements may vary with the sensing tasks, the way semantics manifests in an ISAC channel exhibits a common hierarchical structure. A generic hierarchy that aligns semantic levels with corresponding channel characteristics is detailed below and summarized in Fig.~1.

\textbf{Component-level semantics:}
In many ISAC scenarios especially at high carrier frequencies and with large bandwidth/arrays, an ``object'' is not perceived as a monolithic scatterer but as a composition of salient parts (e.g., vehicle wheels, and UAV wings). Component-level semantics provides a direct handle on the path-level channel formation and is manifested in the coupling among per-path amplitude, delay, and angles, along with the associated micro-Doppler patterns.

\textbf{Object-level semantics:}
Semantics describes the fundamental entities in the environment, including (i) sensing targets with attributes such as type and behavior, and (ii) background elements characterized by coarse geometry and material-related properties. Object-level semantics primarily shapes the intra-cluster channel characteristics, i.e., the amplitude, delay, angle, and Doppler features within a cluster associated with a specific target or a segment of the background. 

\textbf{Scene-level semantics:}
Semantics describes how objects relate to each other and to the environment (e.g., spatial arrangements, relative positions, and blockage relations). These relationships manifest as inter-cluster characteristics within a snapshot, such as the relative placement of clusters in power-delay-angle domains, multi-bounce and the shadowing effects induced by occlusions. 

\textbf{Intent-level semantics:}
As the highest-level environmental semantics, intent-level semantics describes meaningful events that unfold over time. Accordingly, intent-level semantics is represented by the temporal evolution of channel characteristics across snapshots. For example, a potential vehicle collision can be encoded by two clusters whose trajectories converge in the delay-angle domain while their Doppler shifts indicate decreasing separation. Therefore, intent-level semantics is captured through patterns and trajectories in channel dynamics. 

\begin{table*}[ht!]
\centering
\caption{Comparisons among ISAC channel models.}
\includegraphics[width=0.95\textwidth]{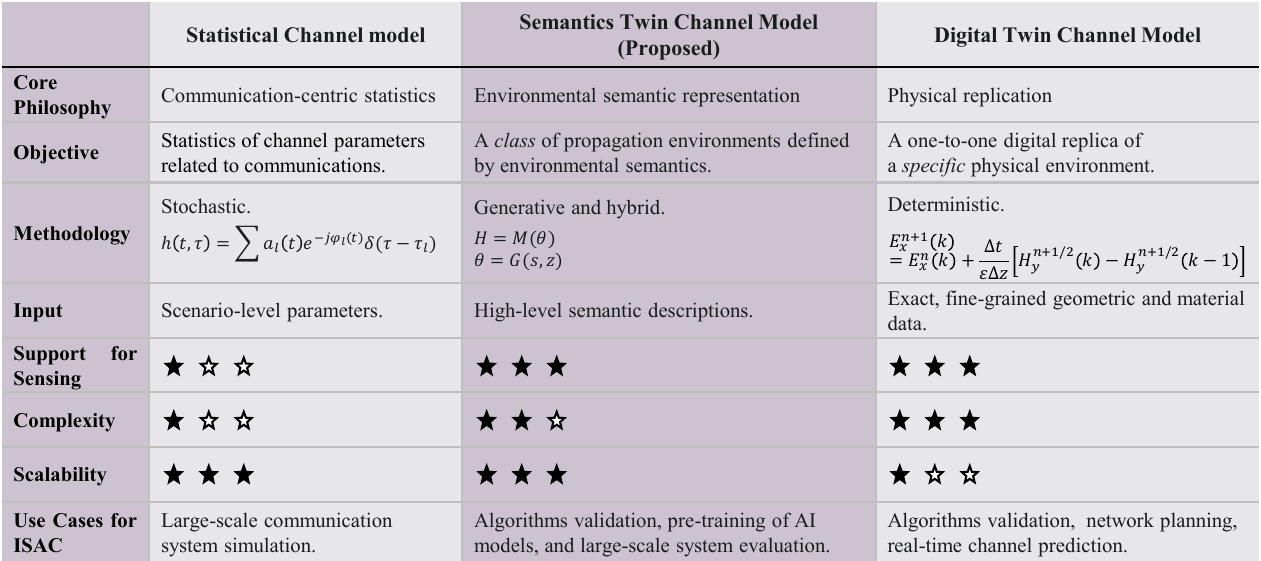}
\label{tab:usage}
\end{table*}

\subsection{Essentiality to ISAC Channel Modeling}
Environmental semantics is the kernel to ISAC channel modeling because it provides the missing ``common language'' that unifies communication- and sensing-oriented channel descriptions. On the sensing side, environmental semantics is the ultimate inference target so that a channel model that does not preserve semantics cannot faithfully support sensing design or evaluation. More fundamentally, semantics is not an add-on annotation to channel responses. The semantic composition of the environment governs the formation of multipath structure and their temporal evolution, which are precisely the channel features that also determine communication performance. Therefore, environmental semantics captures the essence of both ``what the channel means'' for sensing and ``how the channel behaves'' for communication, making it a core design principle in ISAC channel modeling.

Environmental semantics further provides concrete modeling guidance for the trade-off between accuracy and complexity of ISAC channel models, by clarifying what information must be preserved and what can be safely abstracted. On one end, conventional statistical channel models deliberately compress the environment into a small set of communication-oriented statistics. This compression is efficient, but also intrinsically insensitive to many semantically important factors. Consequently, environments with substantially different semantic content can appear statistically similar under a purely statistical channel model. On the other end, deterministic modeling, e.g. ray tracing and computational electro-magnetics, can preserve rich semantic and physical detail, but at the cost of requiring fine-grained environmental inputs and incurring prohibitive computational burden, which restricts scalability across diverse ISAC deployments. 
\subsection{Semantics-oriented ISAC Channel Model}
The discussion of environmental semantics and ISAC channel modeling reveals a fundamental gap in existing ISAC channel modeling. This gap calls for a \emph{semantics-oriented} ISAC channel model, where the key principle is to represent the environment at the semantic granularity that is sufficient to preserve the channel structures that matter for both sensing and communication, while abstracting away physical details that do not alter the semantics of interest. Under this principle, environmental semantics becomes a criterion for model quality and a lever for building controllable representations that generalize and scale across diverse scenarios. The required capabilities of a semantics-oriented ISAC channel model is summarized as below:

\begin{itemize}
    \item \textbf{Conditionable}: The model takes environmental semantics as explicit conditions and preserve a structured semantics-to-channel mapping, so that different semantic conditions induce systematically different channel structures retaining semantic distinguishability.
    \item \textbf{Scalable:} The model efficiently generates a large number of semantically consistent channel realizations at bounded computational cost and without requiring exhaustive site-specific reconstructions. This scalability is essential to represent semantic diversity.
    \item \textbf{Quantifiable:} The model provides quantitative criteria to assess whether generated channels remain faithful to the conditioning semantics, called semantic fidelity. This guarantees semantic consistency of the generated channels and enables model calibration and evaluation.
    \item \textbf{Physically constrained:} The generated channels are physically realizable and comply with fundamental electromagnetic propagation principles. This includes basic constraints such as physically meaningful coupling among amplitude, delay, and angle induced by geometry.
\end{itemize}

\begin{figure*}[ht!]
\centering
\includegraphics[width=0.95\textwidth]{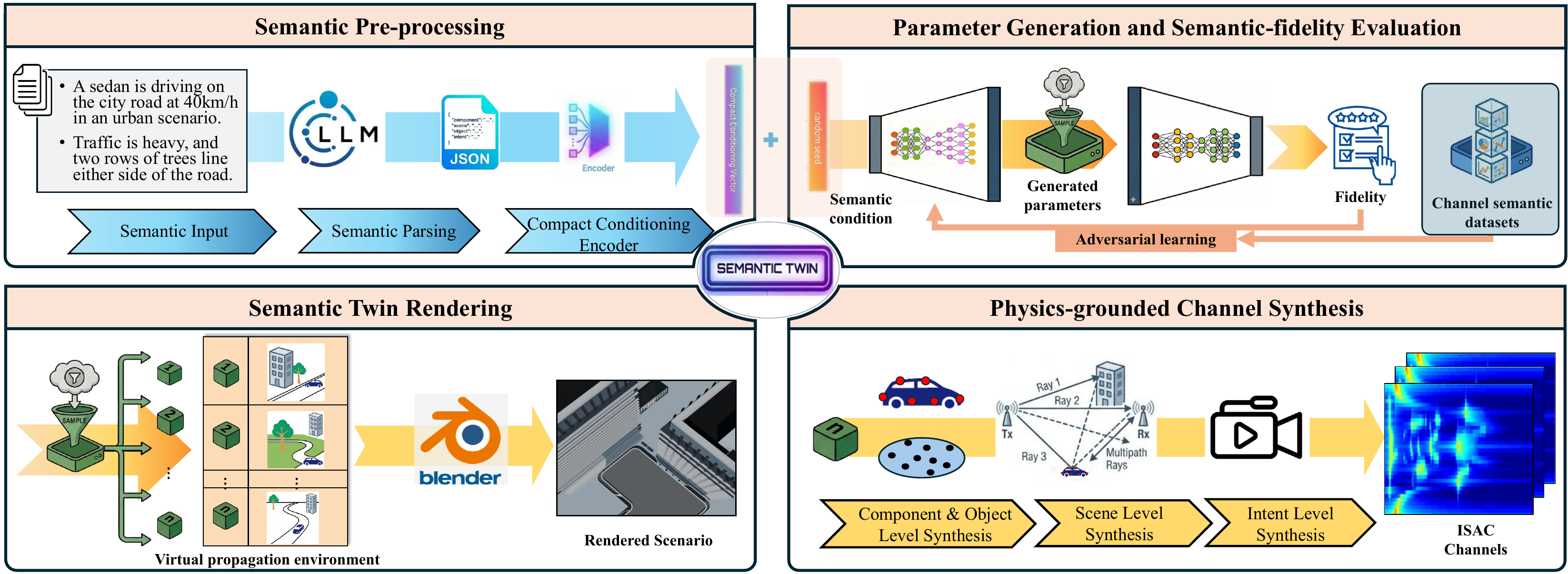}
\caption{Generative AI enabled semantic twin model architecture. }
\label{tab:structure}
\end{figure*}

\section{Semantic Twin Channel Model for ISAC}
In this section, we propose a semantic twin channel model for ISAC which is oriented to the environment semantics and enabled by generative artificial intelligence.
\subsection{Definition}
STCM denotes a environmental-semantics-conditioned, generative channel modeling paradigm for ISAC. Given a semantic description of a scenario, the model produces an ensemble of channel realizations that are representative of a class of environments sharing the same semantics. 
While a digital twin aims to replicate one particular physical site, a semantic twin targets the distribution of channels that can plausibly arise under a given environmental semantic condition, allowing realizations that preserve the same meaning but differ in fine-grained details~\cite{zjh-ChannelGPT}. It is also distinct from purely statistical channel models that compress the environment into a few communication-centric statistics. Instead, it is designed to retain the semantic-to-channel mapping structures that are essential to ISAC.

 By conditioning on semantics and generating many physically realizable channel instances efficiently, a STCM supports large-scale system simulation, robust benchmarking under semantic diversity, and data generation for learning-based ISAC components. These goals naturally motivate a generative-AI-enabled realization, where semantic inputs guide the generation of channel parameters under embedded physical constraints, as developed in the following subsections. The detailed comparisons among statistical, digital twin and proposed STCM is summarized in Table.~I. Note that $M(\theta)$ denotes a physics-grounded channel synthesizer with the parameter set $\theta$, and $G(s,z)$ is a generative model with encoded semantic input $s$ and random vector $z$.

\subsection{Generative AI-enabled Architecture}
We realize the semantic twin channel model through a generative-AI-enabled, semantics-conditioned pipeline that is self-learned from high quality datasets. As illustrated in Fig.~\ref{tab:structure}, the workflow consists of four stages: (i) semantic parsing and encoding, which maps multi-level environmental semantics into a compact conditioning representation; (ii) physics-based channel synthesis, which transforms these parameters into high-dimensional channel realizations while enforcing electromagnetic realizability; (iii) conditional parameter generation, which samples hybrid-model parameters under the semantic condition to yield diverse yet semantically consistent instances;  and (iv) semantic-fidelity evaluation, which quantitatively assesses consistency between the generated outputs and the conditioning semantics, thereby closing the learning loop. In our implementation, stages (i), (iii), and (iv) are realized by neural networks and trained in a measurement-driven, adversarial self-learning manner, whereas stage (ii) provides the physics-grounded synthesis that anchors the generated channels to fundamental propagation principles.

\subsubsection{Environmental Semantic Parsing and Encoding}
Environmental semantics is the conditioning input of a STCM. To make semantics actionable for generation, we adopt a two-step pipeline consisting of LLM-based semantic parsing and compact encoding. In the first step, the user-provided environmental semantics is treated as the \emph{user prompt} to an LLM. We further design a dedicated \emph{system prompt} that specifies (i) the four-level semantics taxonomy established in Section II, (ii) the required fields and naming conventions at each level, and (iii) a strict output schema, so that the LLM consistently parses raw context descriptions into a structured semantic specification in a predefined JSON format. This hierarchical parsing preserves component and object semantics describe the constituents and attributes of entities, scene semantics captures relations and occlusion/visibility context, and intent semantics summarizes time-extended events that drive channel evolution.
\par In the second step, we transform the parsed JSON semantics into a compact conditioning code $s$ using an unsupervised variational autoencoder (VAE). The motivation is that several semantic fields produced in Step 1 are naturally discrete, sparse, and high-dimensional if represented naively. Typical examples include target categories or sub-types, part/component tokens, material/texture labels. Such representations are inefficient for learning and tend to yield sparse conditioning signals. Concretely, VAE converts the JSON into a dense, low-dimensional latent code $s$ that preserves semantic content while providing smoothness and continuity in the conditioning space, improving training stability and enabling controllable semantic interpolation and intra-class diversity.
\begin{figure*}[ht!]
\centering
\includegraphics[width=0.8\textwidth]{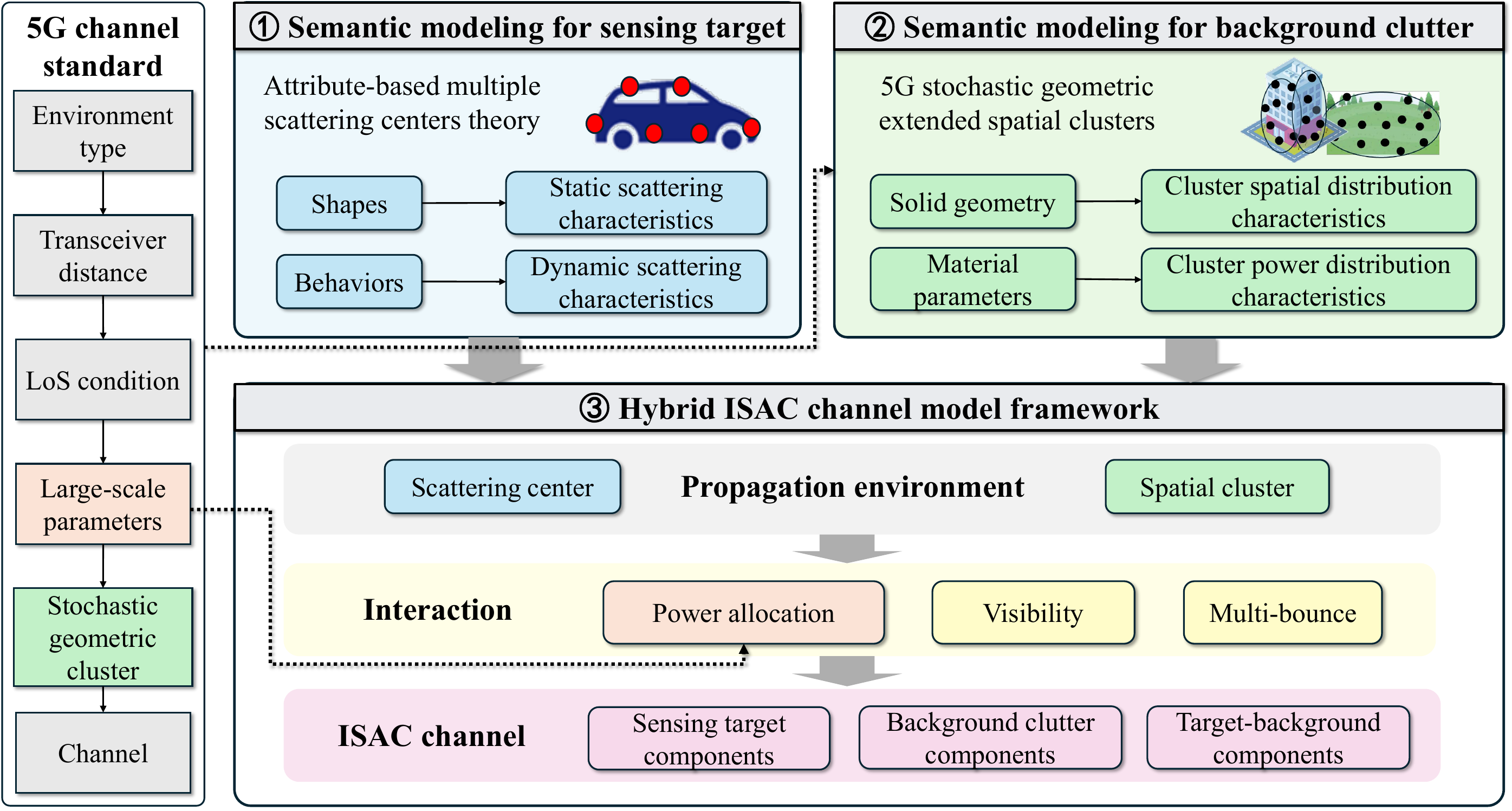}
\caption{Hybrid ISAC channel synsthesizer with environment semantics.}
\label{tab:usage}
\vspace{-0.5cm}
\end{figure*}
\subsubsection{Physics-Grounded Channel Synthesis}
To anchor the semantic twin to real-world propagation, we do not directly synthesize high-dimensional channel responses from a black-box generator. Instead, we introduce a hybrid, physics-grounded ISAC channel synthesizer whose parameters are generated under semantic conditioning and whose channel realizations are produced through physically consistent propagation mechanisms. It decomposes the channel into target scattering, environmental clutter, and target–clutter interactions, so that each semantic level has a clear and controllable entry point into the channel formation process~\cite{lch-rt}.

The hybrid synthesizer combines complementary building blocks that naturally align with the semantic hierarchy. Target-related components capture object- and component-level scattering signatures. Clutter/background components describe the environment-induced multipath structure in a parsimonious, statistically structured form to reflect scene-level context. Interaction components encode visibility, occlusion, and multi-bounce mechanisms that couple targets and clutter, which is essential for reproducing semantically meaningful inter-cluster organization and time variation. With this synthesizer, physical constraints such as causality, kinematics-consistent Doppler behavior, and physically meaningful parameter coupling are enforced by construction, while the learnable generator focuses on producing semantically consistent parameter instances.
The architecture of this model is composed of three primary building blocks:
\begin{itemize}
    \item \textit{Modeling of Targets:}
we adopt the attributed scattering center (ASC) model, which describes a physical object as a set of independent scattering points~\cite{cy-MSC}. The attributed parameters of these points are directly tied to the target’s geometry and material characteristics, thereby establishing an explicit link between object/component-level semantics and the target’s path- and cluster-level scattering signature. The behavioral semantics is modeled through micro-Doppler modeling~\cite{zzy-gesture}. The bulk Doppler shift induced by translational motion, internal or periodic movements generate distinctive Doppler patterns.

    \item  \textit{Modeling of Clutter:}
clutter comprises semantically meaningful objects, e.g., buildings, roadsides, and vegetation, and their EM parameters and spatial organization, which jointly shape the multipath structure. We therefore employ a geometric-based stochastic method aligned with 3GPP-compatible channel standardizations, representing clutter as a collection of scattering clusters distributed in 3D space~\cite{hrs-ISAC}. Clutter semantics is captured by the parameters governing the cluster process, such as the distributions of cluster locations, densities, angular/delay spreads, and powers. 

    \item  \textit{Interactions among targets and clutter:}
after obtaining the attributed scattering centers for targets and the cluster-based representation for the background environment, we account for multipaths arising from their interactions. Using simplified geometrical optics, we model visibility and shadowing effects as well as multi-bounce mechanisms in which signals reflect from both target and environment. These interaction paths are essential for reproducing semantically meaningful coupling between target and clutter, including changes in inter-cluster structure and the time-varying connectivity of multipath components as objects move and events unfold.
\end{itemize}

\subsubsection{Parameter generation and Semantic Fidelity evaluation}
Given the semantic conditioning code $s$, the STCM samples the hybrid channel model parameter set to generate high-dimensional channel responses. Specifically, a generator produces $\boldsymbol{\theta}=G(s,z)$ from the semantic code $s$ and a random latent vector $z$, where sampling $z$ enables diverse yet semantically consistent parameter realizations. The physics-based synthesizer then maps $\boldsymbol{\theta}$ to channel responses under electromagnetic constraints.

To quantify whether the generated parameters are faithful to the conditioning semantics, we define \textit{environmental semantic fidelity} as a distributional metric: for a fixed $s$, it measures the discrepancy between the generated conditional distribution of $\boldsymbol{\theta}$ and the measurement-derived conditional distribution. We instantiate this discrepancy using the Wasserstein distance, interpreted as the minimum transport cost to align the generated and real parameter distributions under the same semantic context.

As this fidelity is distributional and difficult to compute directly in high dimensions, we adopt a generative AI solution with two coupled neural networks trained in an adversarial, data-driven self-learning manner~\cite{HYT}:
\begin{itemize}
    \item \textit{Semantic Fidelity Evaluator:}
The evaluator (critic) takes $(s,\boldsymbol{\theta})$ as input and outputs a scalar score that reflects the semantic fidelity of $\boldsymbol{\theta}$ under condition $s$. Functionally, it serves as a learnable proxy for the Wasserstein-based discrepancy and provides the quantitative metric to compare generated samples against realistic references.

\item \textit{Channel Parameter Generator:}
The generator takes $(s,z)$ as input and synthesizes $\boldsymbol{\theta}$, i.e., a complete parameter instance of the hybrid ISAC channel model. By sampling $z$, the generator produces an ensemble of parameter realizations under the same semantics, enabling scalable generation with controlled diversity.
\end{itemize}

Training proceeds by alternating optimization of the evaluator and the generator in a minimax (adversarial) fashion: the evaluator improves its ability to distinguish measurement-consistent parameters from generated ones under the same semantic condition, while the generator is driven to produce parameters that maximize semantic fidelity. As a result, the learned conditional generator produces parameter ensembles that are both representative of measured propagation and aligned with the input semantics, and the downstream physics-based synthesizer converts them into physically realizable channel responses.
\subsubsection{Semantic Twin Rendering}
To make the parsed semantics tangible and to provide an explicit propagation space for the semantic twin channel model, we introduce a \emph{Semantic Twin Rendering} module. This module takes the structured JSON output from Step 1 as input and instantiates a corresponding semantic scene in Blender. The rendered scene serves as a visual surrogate of the propagation environment under the given semantic condition, enabling both human inspection and downstream visual conditioning. Its outputs include multi-view, time-indexed scene images under specified camera viewpoints for different snapshots as well as visualizations of multipath propagation.

\section{Case Study}
\subsection{Model Training}

To instantiate the proposed semantics-oriented STCM, we train the generative modules in a data-driven manner. For target-related training data, we construct the target dataset via large-scale full-wave EM simulations on the CST platform~\cite{HYT}. For each target, the object is observed from numerous uniformly distributed directions at 10~GHz. We then extract an MSC attribute set with ten dominant scattering centers per direction using a 3D-ESPRIT procedure. The resulting realistic corpus contains 2448 MSC sets for vehicle and UAV targets, obtained from 19,584,000 EM simulations. Second, we generate the clutter training datasets using synthetic channels produced by an extension of 3GPP TR 38.901 model, where the environment-induced multipath is represented by 3D scattering clusters~\cite{lch-rt}. This dataset provides abundant and diverse samples under different environmental semantics conditions. 
\begin{figure}[htb]
\centering
\begin{subfigure}[b]{0.7\linewidth} 
    \includegraphics[width=\linewidth]{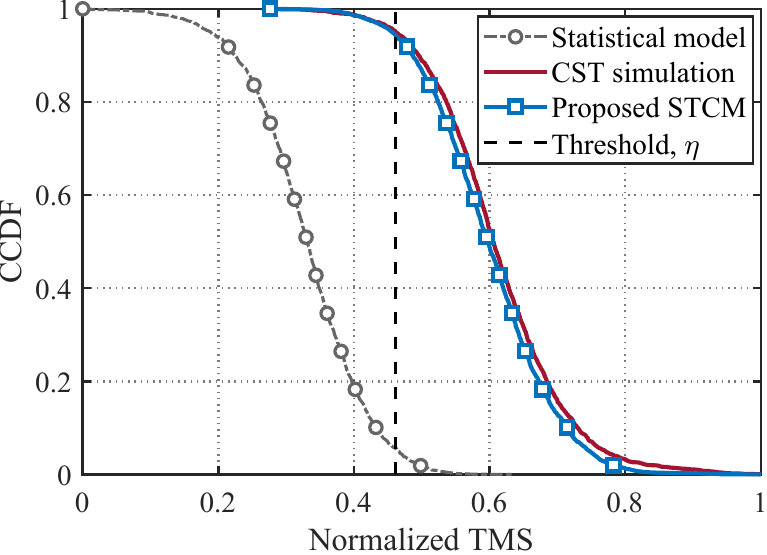}
    \caption{CCDF of TMS for vehicle and UAV targets.}
\end{subfigure}

\begin{subfigure}[b]{0.7\linewidth} 
    \includegraphics[width=\linewidth]{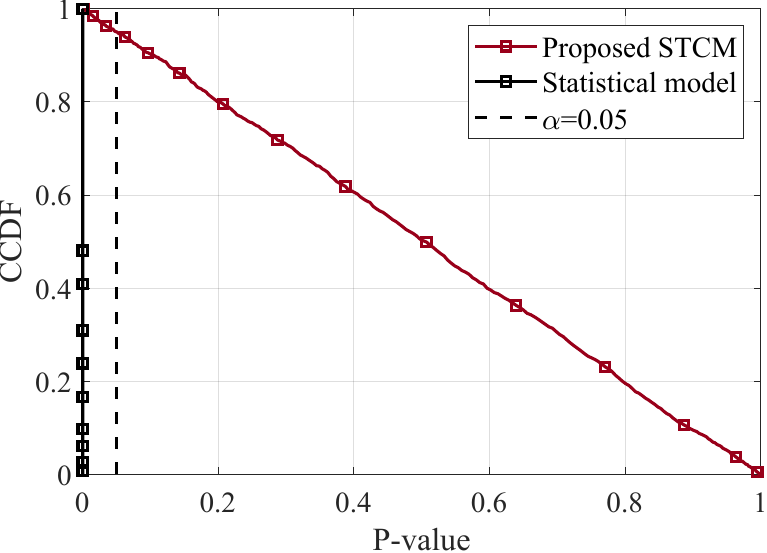} 
    \caption{CCDF of P-values.}
\end{subfigure}
\caption{Experiment results for (a) single-observation identification (b) collaborative identification.}
\label{fig:results}
\vspace{-0.5cm}
\end{figure}
\subsection{Model Evaluation}
To evaluate whether the trained semantic twin preserves environmental semantics in a task-relevant manner, we adopt \textit{target identification} as an sensing task. We quantify identification quality through a target matching score (TMS) computed between a generated channel containing targets and a reference library. As a baseline, we include a statistical target model that matches coarse statistics of MSC attributes. In this article, we report two task-level validations. The first evaluates single-observation identification using the complementary CDF (CCDF) of TMS, where a target is considered identifiable when its TMS exceeds a threshold. The second evaluates collaborative identification using the distribution of K–S test p-values across multiple observing base stations, which probes whether the multi-view TMS pattern produced by a channel model is consistent with that of realistic targets.
\begin{itemize}
    \item \textit{Single-observation identification:} Fig.~\ref{fig:results}(a) shows the CCDF of normalized TMS for both vehicle and UAV targets, comparing realistic references, the proposed model, and the statistical baseline. Following the protocol, the decision threshold is chosen such that the realistic reference achieves a 95\% exceedance probability. Under the same threshold, the proposed model yields exceedance probabilities of 94.24\%, whereas the statistical baseline collapses to 5.56\%.
    \item \textit{Collaborative identification:} Fig.~\ref{fig:results}(b) plots the p-value distribution evaluated from 10 observing base stations. The proposed STCM achieves 95.03\% of p-values exceeding a significance level, whereas the statistical baseline yields p-values approaching zero, corresponding to an almost always rejection outcome. 
\end{itemize}
Across both single-observation and collaborative identification, the proposed STCM demonstrates that semantics-preserving generation is achievable at the task level. It reproduces the discriminative target signatures required for identification and remains consistent under multi-view collaborative sensing, where the statistical model breaks down.
\vspace{-0.1cm}

\section{Challenges and Future directions}
\subsection{Channel Semantics Dataset}
Channel Semantics Dataset. A key bottleneck for scaling semantics-oriented ISAC channel modeling is the lack of a channel semantics dataset that couples real-world channel measurements with multi-level semantic annotations~\cite{molisch-measurement}. A broadly deployable semantic twin ultimately requires data that jointly captures (i) hierarchical semantics, (ii) synchronized sensing–communication observations, and (iii) diverse conditions spanning frequency bands, bandwidths, array apertures, mobility, and occlusion dynamics. The central challenge is not only collecting channel data at scale, but also producing reliable semantic labels and semi-automated annotation so that the conditional generator can learn a faithful semantics-to-parameter distribution rather than overfitting to narrow proxies.

\vspace{-0.35cm}
\subsection{Multi-modal STCM}
A natural evolution of the proposed STCM is to make it multi-modal, e.g., Lidar, camera and infrared devices~\cite{3dgs}. First, multi-modal sensing can assist semantic annotation by providing reliable cues for objects, relations, and events, thereby improving the efficiency and quality of channel semantics datasets and reducing the cost of large-scale labeling. Second, multi-modal observations provide richer and less ambiguous semantic descriptions than symbolic labels alone. Geometry, material proxies, visibility/occlusion, and kinematic context extracted from different modalities can be fused into a more informative conditioning input, which improves semantic fidelity and generalization of the generated channels. Third, multi-modal outputs naturally enable broader ISAC application scenarios, i.e., collaborative multi-sensing.

\vspace{-0.35cm}
\subsection{Online Adaptation and Continual Learning}
A STCM learns a semantics-conditioned channel distribution, and this distribution drift over time even when the semantic appears unchanged. For example, ISAC environments exhibit such a drift that an ``urban street'' can change gradually as building materials shift toward glass facades, traffic density increases, and roadside infrastructure proliferates, thereby altering the multipath structure and interaction patterns associated with the same semantic condition and degrading semantic fidelity. This motivates online adaptation and continual learning that detect drift using fidelity signals, recalibrate the model with limited new measurements, and absorb new semantic patterns without catastrophic forgetting.

\section{Conclusion}
In this article, we positioned environmental semantics as the unifying abstraction for ISAC channel modeling and showed that semantics can be systematically linked to observable channel structures through a hierarchical semantics-to-channel mapping. Based on this perspective, we advocated a semantics-oriented modeling principle that balances accuracy and complexity by preserving semantics-relevant channel structure while avoiding unnecessary physical detail. To realize this principle, we introduced a generative AI–empowered STCM that produces an ensemble of physically plausible channel realizations conditioned on a semantic description. A physics-grounded hybrid synthesizer and a distributional semantic-fidelity criterion further enable controllable generation and quantitative validation. The case-study results suggest that semantics-preserving channel generation is achievable and scalable, supporting reproducible ISAC simulation, dataset generation, and benchmarking, and providing a practical pathway toward future ISAC design and standardization.

\vfill


\begin{thebibliography}{1}

\bibliographystyle{IEEEtran}

\bibitem{itu}
{\textit{Framework and overall objectives of the future development of IMT for 2030 and beyond}}, Rec. ITU-R M.2160, International Telecommunication Union, Geneva, Switzerland, Nov. 2023. [Online]. Available: https://www.itu.int/rec/R-REC-M.2160/en
\bibitem{ywf}
W. Yang et al., ``Integrated sensing and communication channel modeling and measurements: Requirements and methodologies toward 6G standardization,'' {\textit{IEEE Vehicular Technology Magazine}}, vol. 19, no. 2, pp. 22-30, June 2024.
\bibitem{zjh-ISAC}
J. Zhang et al., ``Integrated sensing and communication channel: Measurements, characteristics, and modeling,'' {\textit{IEEE Communications Magazine}}, vol. 62, no. 6, pp. 98-104, June 2024.
\bibitem{gk-ISAC}
T. Liu et al., ``6G integrated sensing and communications channel modeling: Challenges and opportunities,'' {\textit{IEEE Vehicular Technology Magazine}}, vol. 19, no. 2, pp. 31-40, June 2024.
\bibitem{channel-semantics}
Z. Gao, S. Liu, Y. Su, Z. Li and D. Zheng, ``Hybrid knowledge-data driven channel semantic acquisition and beamforming for cell-free massive MIMO,'' \textit{IEEE Journal of Selected Topics in Signal Processing}, vol. 17, no. 5, pp. 964-979, Sept. 2023

\bibitem{zzy-cs}
Z. Zhang at al., ``Channel semantic characterization for integrated sensing and communication scenarios: From measurements to modeling.'' \textit{arXiv preprint} arXiv:2503.01383, 2025.

\bibitem{zjh-ChannelGPT}
L. Yu at al, ``ChannelGPT: A large model toward real-world channel foundation model for 6G environment intelligence communication,'' \textit{IEEE Communications Magazine}, vol. 63, no. 10, pp. 68-74, Oct. 2025.

\bibitem{lch-rt}
C. Luo et al, ``Channel modeling framework for both communications and bistatic sensing under 3GPP standard,'' {\textit{IEEE Journal of Selected Areas in Sensors}}, vol. 1, pp. 166-176, 2024.

\bibitem{wcx-ISAC}
R. Yang, C. -X. Wang, J. Huang, E. -H. M. Aggoune and Y. Hao, ``A novel 6G ISAC channel model combining forward and backward scattering,'' {\textit{IEEE Transactions on Wireless Communications}}, vol. 22, no. 11, pp. 8050-8065, Nov. 2023.

\bibitem{cy-MSC}
Y. Chen et al., ``Multi-scattering centers extraction and modeling for ISAC channel modeling,'' in {\textit{Proc. 18th European Conference on Antennas and Propagation}}, Glasgow, United Kingdom, 2024, pp. 1-5.
\bibitem{zzy-gesture}
Z. Zhang et al., ``Deep learning-based human gesture channel modeling for integrated sensing and communication scenarios,'' \textit{IEEE Transactions on Antennas and Propagation}, early access, 2025.
\bibitem{hrs-ISAC}
Z. Zhang et al., ``A general channel model for integrated sensing and communication scenarios,'' {\textit{IEEE Communications Magazine}}, vol. 61, no. 5, pp. 68-74, May 2023.
\bibitem{HYT}
Y. Hu and Yi Chen, ``Sensing-task-oriented generative channel model for 6G integrated sensing and communications,'' in \textit{Proc. of IEEE Wireless Communications and Networking Conference (WCNC)}, 2026.
\bibitem{molisch-measurement}
H. Hammoud et al., ``A novel low-cost channel sounder for double-directionally resolved measurements in the mmWave band,'' {\textit{IEEE Transactions on Wireless Communications}},  vol. 24, no. 1, pp. 340-354, Jan. 2025.

\bibitem{3dgs}
B. Fei, J. Xu, R. Zhang, Q. Zhou, W. Yang and Y. He, ``3D gaussian splatting as new era: A survey,'' {\textit{IEEE Transactions on Visualization and Computer Graphics}}, vol. 31, no. 8, pp. 4429-4449, Aug. 2025.



\end{thebibliography}
\end{document}